\documentclass[prl,amssymb,amsmath,twocolumn,floatfix,letterpaper]{revtex4}
\usepackage{hyperref}
\usepackage{graphicx}
\begin{document}

\title{Boundary hopping and the mobility edge in the Anderson model
 in three dimensions}

\author{ Viktor Z. Cerovski }
\affiliation{ Institut f\"ur Physik, Technische Universit\"at, D-09107
Chemnitz, Germany. } 

\begin{abstract}
It is shown, using high-precision numerical simulations, that the mobility edge
of the 3d Anderson model depends on the boundary hopping term $t$ in the
infinite size limit.  The critical exponent $\nu$ is independent of it.  The
renormalized localization length at the critical point is also found to depend
on $t$ but not on the distribution of on-site energies for box and Lorentzian
distributions.  Implications of results for the description of the transition
in terms of a local order-parameter are discussed.  
\end{abstract}

\date{May 17, 2006}

\maketitle

Since the discovery of the Anderson localization~\cite{Anderson58}, one of the
central problems of the physics of disordered materials was to understand the
metal-insulator transition induced by the localization of electronic states.
The scaling theory of localization~\cite{Abrahams79} proposed that the
disorder-averaged dimensionless conductance $\langle g \rangle$ is the only
relevant scaling parameter governing the transition, and that there is only one
fixed point under the renormalization-group transformation (RG) in 1d and 2d
corresponding to the insulating phase, with 2 being the lower critical
dimension of the transition, while in 3d there are three such points and a
continuous localization-delocalization transition characterized by a critical
exponent $\nu$.

The discovery of the universal conductance fluctuations~\cite{Mesoscopic91}
showed that $g$ is not a self-averaging quantity, and therefore scaling of its
whole distribution has to be studied. Nonetheless, there is a compelling
evidence that the scaling properties of the distribution itself in the critical
region of the transition are still governed by a single-parameter~%
\cite{Shapiro86}.

The transition has a field-theoretical description within the super-symmetric
nonlinear $\sigma$-model~\cite{Efetov83} (NLS), but technical difficulties with
the calculation of $\nu$ from the $2+\epsilon$ expansion do not yet permit
accurate determination of $\nu$, despite significant theoretical
progress~\cite{Hikami81}.  
$\nu$ is currently most accurately determined numerically, especially {\it via}
the transfer-matrix method (TMM)~%
\cite{Pichard81,MacKinnon81,MacKinnon94,Slevin99}, also
used in this work and described below. TMM gives the
phase diagram of the transition, by calculating the critical disorder
strength, $W_c$, at which a critical state appears at the band center ($E=0$).
For $W<W_c$ there is a band of extended states within $|E|<E_c(W)$. The 
mobility edges $W_c$ and $E_c(W)$ depend on the distribution of impurity 
energies~\cite{Bulka85}.

The subtlety of the Anderson transition is that natural choices for an order
parameter, such as the density of states or the $Q$ fields, that are conjugated
to the symmetry breaking field in an effective field-theoretical description of
the transition within NLS, behave analytically in the vicinity of the
transition~\cite{Wegner76,Efetov83}.  Nevertheless, a description of the
transition in terms of an order parameter {\it function} (OPF) has been
proposed~\cite{Zirnbauer86} and shown to be related in a deep way to the
distribution of the local density of states~\cite{Mirlin94}.

Another surprising feature of the transition is the influence of boundary
conditions (b.c) in the critical region on several important quantities:
critical distribution of $g$~\cite{Soukoulis99comment},
critical level spacing~\cite{Shklovskii93,Braun98}, critical renormalized
localization length $\Lambda_c$~\cite{Slevin00} (that is also studied in this
work) as well as on the $\beta$-function of $\langle g\rangle$ in diffusive
approximation~\cite{Braun01}.  These are generally understood in terms of the
dependence of the quantized diffusion modes on the choice of
b.c~\cite{Braun98}, the sample geometry~\cite{Potempa98} and
topology~\cite{Kravtsov99}. The case of general boundary conditions was
recently considered in Ref.~\cite{Evangelou05}.

To further clarify the role of b.c and surface effects 
on the properties of the localization-delocalization transition, a model is
proposed here that corresponds to the Anderson model with boundary hopping
terms taken to be continuous variables. In this model one expects diffusion
modes to be quantized in a quantitatively different way than in the models
with periodic or open b.c.  The principal finding of this Letter is that the
mobility edge of the Anderson model in 3d depends on the boundary hopping term.

The studied Hamiltonian is:
\begin{eqnarray*}\label{eq:Hamiltonian}
H & = &  \sum_i\epsilon_i|i\rangle\langle i| \;+\; t_b
  \!\!\!\!\!\!\!\!\!\!\!\!\!\!\!
  \sum_{|i_x-j_x|+|i_y-j_y|+|i_z-j_z|=1}
  \!\!\!\!\!\!\!\!\!\!\!\!\!\!\!
       \left(|i\rangle\langle j| + |j\rangle\langle i|\right) \\
  & + & t_x\!\!\!\sum_{\begin{subarray}{c} i_x=1,\,j_x=L_x\\
  i_y=j_y,\,i_z=j_z\end{subarray}}
     \!\!\!\left(|i\rangle\langle j|+|j\rangle\langle i|\right) \\
  & + & t_y\!\!\!\sum_{\begin{subarray}{c} i_y=1,\,j_y=L_y\\
  i_x=j_x,\,i_z=j_z\end{subarray}}
     \!\!\!\left(|i\rangle\langle j|+|j\rangle\langle i|\right)\\
  & + & t_z\!\!\!\sum_{\begin{subarray}{c} i_x=1,\,j_x=L_x\\
  i_x=j_x,\,i_y=j_y\end{subarray}}
     \!\!\!\left(|i\rangle\langle j|+|j\rangle\langle i|\right),
\end{eqnarray*}
where $i,j$ denote sites of the simple cubic lattice of $L_x\times L_y\times
L_z$ atoms; $|i\rangle$ is the eigenstate of the electron localized at
the $i$-th site with impurity energy $\epsilon_i$, which is an i.i.d random
variable with the probability distribution $P(\epsilon,W)$, here taken to be
box distribution (BD), i.e. $\epsilon_i$ are uniformly distributed in
$[-W/2,W/2)$, and Lorentzian distribution (LD),
$P(\epsilon,W)=W/\pi(W+\epsilon^2)$.  The hopping integral in the bulk, $t_b$,
is set to 1.  By taking $t_{x,y,z}$ to be 0 or 1 one obtains, as special cases,
models with open, periodic and mixed boundary conditions studied in
Ref.~\cite{Braun98,Slevin00}.  

This work studies the geometry of quasi-1d slabs 
$L=L_x=L_y$ and $L_z=M\gg L$, 
(the density of boundary hoppings is thus $L^{-1}$), 
and 
the localization properties of the 3d samples are deduced from the scaling
properties of the smallest Lyapunov exponent when $L\to\infty$.  To this end
the eigenproblem $H\Psi = E\Psi$ is rewritten in terms of the transfer
matrices, 
$\Psi_M = \mathcal{T}_M \Psi_0,\; 
\mathcal{T}_M = \prod_{i=1}^M\left(\begin{array}{cc} H_i - E & -1 \\
                                                         1   & 0
                                   \end{array}\right),$
where $H_i, \Psi_i$ are, respectively, projections of $H$,$\Psi$ on the $i$-th
slice of the slab.  The product converges towards the limiting symplectic
matrix when $M\to\infty$, with corresponding $2M^2$ Lyapunov exponents
$\gamma_i$ being eigenvalues of
$
  \lim_{M\to\infty}\frac{1}{2M}\ln \mathcal{T}_M^\dagger \mathcal{T}_M,
$
which are calculated to a given relative accuracy by multiplication of
a sufficient number of $\mathcal{T}_i$ combined with the Gramm-Schmidt
reorthonormalization~\cite{MacKinnon81,MacKinnon94,Slevin04}.  Fermi energy is
set $E=0$, since the critical states are expected to appear first at the band
center when $W$ is decreased.

The renormalized inverse of the algebraically smallest $\gamma_i$,
$\Lambda\equiv(L\min_i|\gamma_i|)^{-1}$, defines the largest length scale in
the system that is identified with the correlation length, the scaling
law of which near the critical point is~\cite{Cardy96}:
\begin{equation}
 \Lambda(w,L) = F(\psi(w)L^{1/\nu},\phi(w)L^{y}),\; w=\frac{W-W_c}{W_c},
\end{equation}
where $\nu>0$ is the critical exponent, $y<0$ is the irrelevant critical
exponent due to the finite-size correction to scaling and $W_c$ is the critical
disorder strength. Near the critical point $(w=0, L^{-1}=0)$ the scaling
function $F$, the relevant scaling field $\psi$ and the irrelevant scaling
field $\phi$ are expected to be analytic~\cite{Cardy96}, and each of them can
be expanded giving the following class of model functions for the fit:
\begin{eqnarray}
\Lambda(w,L)&=&F_0(\psi(w) L^{1/\nu})+\\
            &+&\sum_{n=1}^{n_I}\phi(w)^n L^{n y}F_n(\psi(w) L^{1/\nu}),
            \label{corrections}\\
F_n(\psi(w) L^{1/\nu})&=&\sum_{m=0}^{k_n} F_{nm}\psi(w)^m L^{m/\nu},\label{F}\\
 \psi(w)&=&\sum_{j=1}^{m_R} \psi_j w^j,\;
 \phi(w) = \sum_{j=0}^{m_I} \phi_j w^j;\label{f}
\end{eqnarray}
where coefficients of expansions are fitting parameters (except for
$F_{01}=F_{10}=1$ because two parameters are not independent~\cite{Slevin99}),
that are determined from the least-$\chi^2$ nonlinear fit.  We generalize the
class of model functions from Ref.~\cite{Slevin99} by allowing each $F_n$ to be
expanded up to $k_n$-th power, where $k_n$ are not necessarily all equal, and
by allowing some of $\phi_j$ to be set to 0 for $j<m_R$, giving $m_R'\le m_R$
coefficients of the expansion of $\psi(w)$.  The total number of fitting 
parameters is thus $N_p = \sum_{n=0}^{n_I}(k_n+1)+m_R'+m_I-1$.

$F_0$ is the one-parameter scaling function, while 
Eq.~(\ref{corrections}) represents corrections to scaling due to the
finite-size effects described by $y$ and $\phi$.
The corrections to scaling were studied in the context of the integer quantum
Hall effect~\cite{Huckestein94,Wang96,Schweitzer05}.  They were
shown~\cite{Huckestein94,MacKinnon94} to be important for the accurate
estimation of $\nu$, while the formulation of Ref.~\cite{Slevin99} allows also
accurate determination of $W_c$. 

The case $t=t_x=t_y$ is studied first.
To determine the dependence on $t$, Lyapunov exponents were calculated and the
scaling analysis done independently for several values of $t$.
A relative accuracy of $0.05\%$ was chosen in all cases studied, and
correspondingly the slabs were up to $4\times 10^7$ sites long.  Values of
$N_p$ fitting parameters $\{\alpha_i\} = \{\nu, y, W_c, F_{mn}, \psi_n,
\phi_n\}$ were determined using the Levenberg-Marquard algorithm.  95\%
confidence intervals of $\{\alpha_i\}$ are determined explicitly, by
calculating projections of the confidence region
$\chi^2(\{\alpha_i\})\le\Delta\chi^2$, where $\Delta\chi^2$ depends on $N_p$
and the confidence level~\cite{NRC2ed}.  This is a nontrivial calculation for
which an efficient algorithm was developed.  

Table~1 summarizes parameters of the simulation, obtained values of $\chi^2$
and the ``quality of fit'' parameter $Q$. It was suggested in
Ref.~\cite{Slevin99} that acceptable fits should have $Q>0.1$, which the
results presented here satisfy.

\begin{table}
\begin{center}
\begin{tabular}{c c c c c c c c | c c}
\hline
\hline
disorder& $t$  &      $W$      &  $L$   &$m_R$&$k_1$&$N_d$&$N_p$&$\chi^2$&$Q$\\
\hline
   box  & 0.0  &   $[15,18]$   &$[5,15]$&  3  &  2  & 671 & 11  &  647  &0.6\\
   box  & 0.1  &$[15.25,17.75]$&$[5,15]$&  3  &  3  & 561 & 12  &  580  &0.2\\
   box  & 0.25 & $[15.5,17.5]$ &$[5,16]$&  1  &  2  & 492 & 10  &  472  &0.6\\
   box  & 0.4  & $[15.5,17.5]$ &$[5,15]$&  1  &  2  & 451 & 10  &  452  &0.3\\
   box  & 0.5  & $[15.5,17.5]$ &$[4,14]$&  2  &  1  & 451 & 10  &  392  &0.95\\
   box  & 0.7  &$[15.25,17.75]$&$[4,13]$&  2  &  2  & 510 & 11  &  484  &0.7\\
   box  & 0.9  & $[15.5,17.5]$ &$[4,13]$&  2  &  2  & 410 & 11  &  397  &0.5\\
   box  & 1.0  & $[15.5,17.5]$ &$[4,15]$&  2  &  0  & 492 &  9  &  450  &0.9\\
\hline
 Lorentz& 0.0  &  $[3.8,4.65]$ &$[5,16]$&  3  &  2  & 516 & 11  &  507  &0.5\\
 Lorentz& 0.25 & $[4.05,4.55]$ &$[5,15]$&  2  &  1  & 561 & 10  &  574  &0.2\\
 Lorentz& 0.9  &  $[4.1,4.5]$  &$[5,14]$&  2  &  0  & 810 &  9  &  784  &0.7\\
\hline
\hline
\end{tabular}
\end{center}
\caption{Input parameters of the fit and values of obtained $\chi^2$ and 
$Q$.  $N_d$ is the number of points, $N_p$ is the number of fitting 
parameters.  In all cases $n_R=k_0=3,n_I=1,m_I=0$.  Parameter $\psi_2$ 
is set to 0 for $t=0,0.1$.
} \label{table1}
\end{table}

\begin{figure}
\begin{center}
\includegraphics*[width=3.25in]{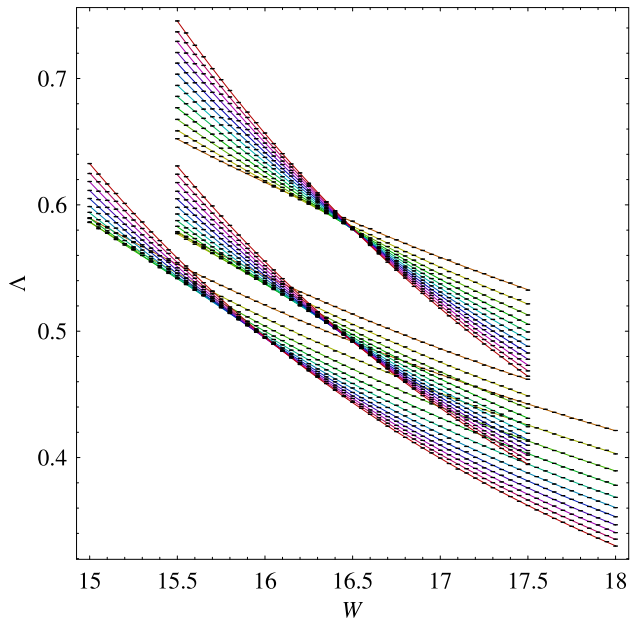}
\end{center}
\caption{(Color online) $\Lambda(W,L)$ calculated for $t=0,0.25,1$, going from
the bottom set of points to the top, with corresponding fits (lines).  In each
set the fastest changing $\Lambda(W)$ corresponds to the largest $L$ studied.
Sets of points for intermediate values of $t$ smoothly interpolate between
the values in the Figure and are omitted for clarity.} \label{fig1}
\end{figure}

The calculated values of $\Lambda(W,L)$ for $t=0, 0.25$ and $1$ are presented
in Fig.~1, together with the lines corresponding to the fits.  The Fig.~shows
an increase of $\Lambda(L)$ with $L$ for smaller values of $W$ and a decrease
for larger $W$, which correspond, respectively, to the delocalized (metallic)
and localized (insulating) behavior.
The same values with the corrections to scaling, Eq.~(\ref{corrections}),
subtracted are shown in Fig.~2, and the critical point is where all curves
cross.  Figure~3 shows the data collapse, where the upper branch corresponds to
the metallic ($w<0$) and the lower branch to the insulating ($w>0$) phase.  

\begin{figure}
\begin{center}
\includegraphics*[width=3.25in]{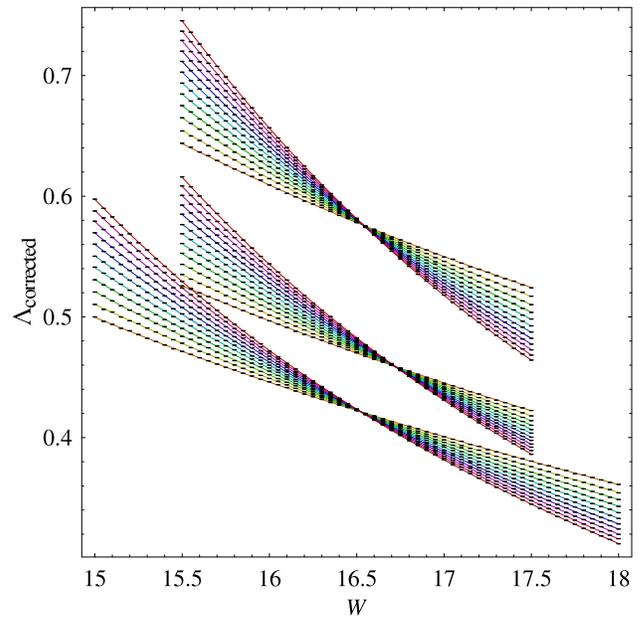}
\end{center}
\caption{(Color online) Same as in Fig.~1 after the subtraction of the
correction to scaling due to the irrelevant field.}
\label{fig2}
\end{figure}

\begin{figure}
\begin{center}
\includegraphics*[width=3.25in]{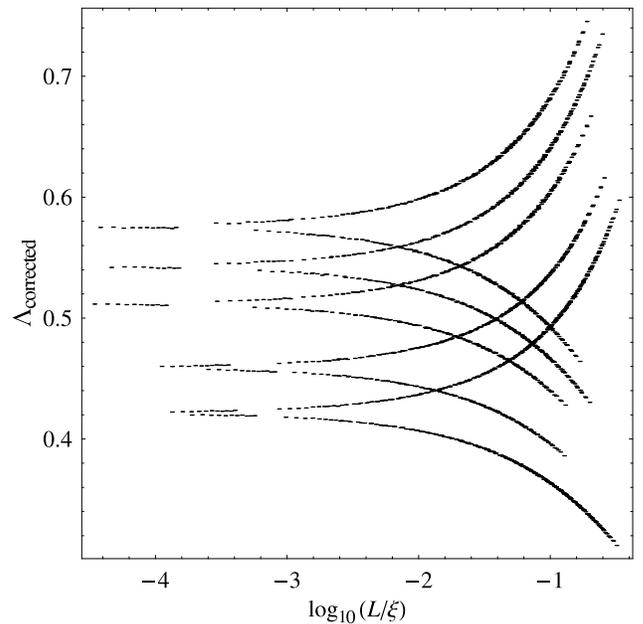}
\end{center}
\caption{Data collapse for $t=0,0.25,0.5,0.7,$ and $1$ (from the
bottom two curves to the top).}
\label{fig3}
\end{figure}

The obtained values of $\nu, \Lambda_c, y,$ and $W_c$ are presented,
respectively, in Fig.~4 and 5. The results in the case of open ($t=0$) and 
periodic ($t=1$) boundaries for BD are in excellent agreement with 
Ref.~\cite{Slevin99,Slevin00}.  They support the universality of $\nu$, 
and non-universal, $t$-dependent, $\Lambda_c$, $W_c$ and $y$.  $\Lambda_c(t)$
is, however, found to be independent of the distribution of $\epsilon$, which 
therefore supports the universality of the {\it function} $\Lambda_c(t)$.

Most remarkably, the results suggest that $W_c(t)$ is not constant,
{\it i.e}, {\it the boundary hopping term $t$ induces a shift of the critical
point of the infinite system}.  To understand this better, we recall that in
the classical systems with short-range interaction and the local
order-parameter, one generally expects surface ordering with a different
critical temperature $T_c$ from that in the bulk, but the $T_c$ of the bulk is
not affected by the presence of boundaries~\cite{Cardy96}.  If these
assumptions are generic for any theory with local order-parameter and
short-range interactions, and having in mind that disorder plays the analogous
role to temperature in classical phase transitions, one arrives 
to the conclusion that there is no such theory in the universality class of the
$H$ studied here for generic values $t_{x,y,z}\ne 0,1$.  It should be noted
that the non-constant $W_c(t)$ found here is not in contradiction with the OPF
description of the transition, since OPF at a given point is defined as an
integral involving values of $Q$-fields at {\it all} other points of the
sample.

\begin{figure}
\begin{center}
\includegraphics*[width=1.1in]{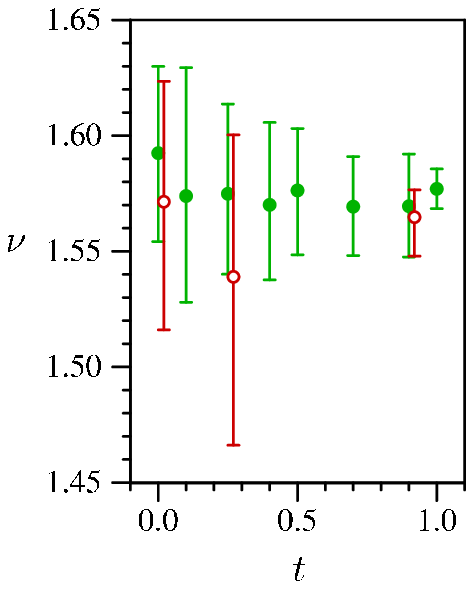}
\includegraphics*[width=1.1in]{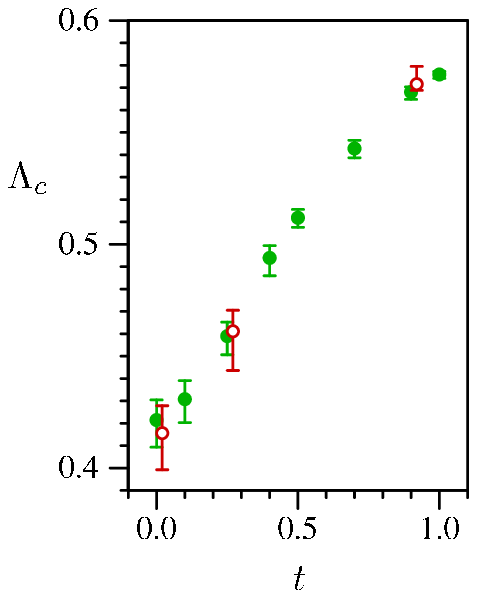}
\includegraphics*[width=1.05in]{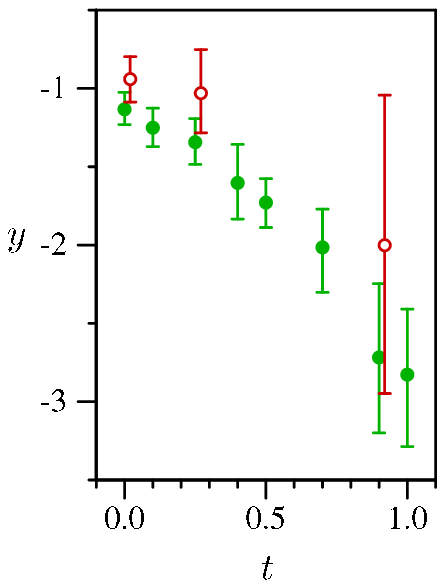}
\end{center}
\caption{
(Color online) Obtained values of the $\nu, \Lambda_c$ and $y$ for box
(filled circles) and Lorentzian (open circles) distributions of disorder.  The
latter are slightly displaced along the $x$-axis for clarity.  
Error bars are $95\%$ confidence intervals. 
}
\label{fig4}
\end{figure}

\begin{figure}
\begin{center}
\includegraphics*[width=1.6in]{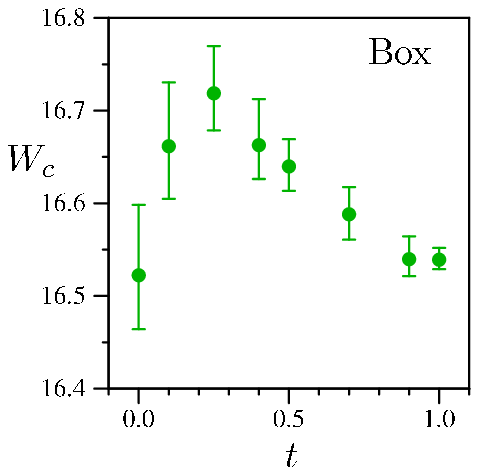}
\includegraphics*[width=1.6in]{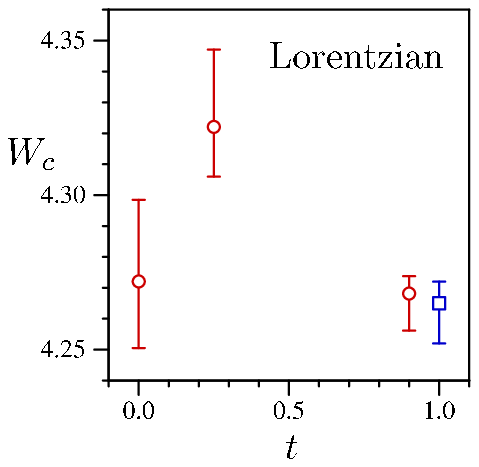}
\end{center}
\caption{
(Color online) Dependence of the critical disorder strength $W_c(t)$ (mobility
edge) for the box (left panel) and Lorentzian (right panel) distribution of
on-site energies, separating the metallic ($W<W_c$) from the insulating
($W>W_c$) phase.  The value $W_c(t=1)$ in the Lorentzian case is from
Ref.~\protect\cite{Slevin99}.  Error bars are $95\%$ confidence intervals.
}
\label{fig5}
\end{figure}

In addition to $H$ with constant $t$, systems where each individual boundary
hopping term, $t_{ij}$, is an i.i.d random variable were studied for three
distributions: $t_{ij}\in [0,1)$, $t_{ij}\in [-1,1)$, and when $t_{ij}$ takes
the values 0 or 1 with the probability 1/2.  In all three cases BD is used for
$\epsilon_i$. The results for $\nu, W_c, \Lambda_c$ and $y$ in all three cases
give values close to those obtained for $t=0.5$, thus suggesting that
randomness in boundary hopping terms can be described by $H$ with an effective
constant $t$.

Furthermore, the $t_x \ne t_y$ case was studied as well for $t_x=0.7,
t_y=0$, when $\Lambda_c \approx 0.47(2)$ was obtained for both BD and LD. Since
they differ from $\Lambda_c(t_x=t_y=0.7)= 0.543_{38}^{47}$,  it seems
reasonable to expect that there is also a universal $\Lambda_c(t_x,t_y)$
function independent of the distribution of on-site energies.

Finally, the $t=-0.1$ case with BD of $\epsilon_i$ was studied to see effects
of the sign of $t$ on critical properties, and results for $\nu, W_c,
\Lambda_c,$ and $y$ are approx.~the same as those obtained for $t=0.1$.  This
is not surprising since it was already found~\cite{Slevin00} that the change
from periodic to antiperiodic b.c. does not affect critical properties.  From
these two results it seems reasonable to infer that
$\Lambda_c(t)\approx\Lambda_c(-t)$.

In conclusion, the Anderson model with the boundary hopping term $t$ is
studied near the localization-delocalization transition in the presence
of the time-reversal symmetry for two distributions of on-site
energies.  After the careful estimation of statistical errors, the
universality of $\nu$ and $\Lambda_c(t)$ is shown, and found that $W_c$
depends of $t$, which is interpreted as the absence of the local
order-parameter description of the transition.  The surprising
deviation of up to $1\%$ of $W_c(t)$ remains to be explained, for
example by a theory that would account for the occurrence of the
maximum of $W_c$ at $t\approx 0.25$.

\end{document}